\documentclass[letterpaper, 10pt, conference]{ieeeconf}
\IEEEoverridecommandlockouts                          
\usepackage{amssymb}
\usepackage{graphicx} 
\usepackage{amsmath}
\usepackage{lineno}
\usepackage{hyperref}
\usepackage{bm}
\usepackage{amsmath}
\usepackage{amsfonts}
\usepackage[dvipsnames]{xcolor}
\usepackage{float}
\usepackage{framed}
\usepackage{makecell}
\usepackage{arydshln}
\usepackage{fancyhdr}
\usepackage{booktabs}
\usepackage{fullpage}
\usepackage[linesnumbered,boxed,commentsnumbered,ruled,vlined,longend]{algorithm2e}
\usepackage{geometry}
\definecolor{darkgreen}{rgb}{0.8,.4,0.4}

\title{\LARGE \textbf{Virtual trajectories for I-24 MOTION: data and tools}}

\author{
Junyi Ji$^{\ast\dagger\P}$, Yanbing Wang$^{\P}$, Derek Gloudemans$^{\P}$, Gergely Zach\'{a}r$^{\P}$, William Barbour$^{\P}$, Daniel B. Work$^{\dagger\P}$
\thanks{$^{\ast}$Corresponding author, Junyi Ji, \url{junyi.ji@vanderbilt.edu}.}\thanks{$\dagger$Department of  Civil and Environmental Engineering, $\P$ Institute for Software Integrated Systems, Vanderbilt University.}
}
\begin{document}
\maketitle

\begin{abstract}
This article introduces a new virtual trajectory dataset derived from the I-24 MOTION INCEPTION  v1.0.0
dataset to address challenges in analyzing large but noisy trajectory datasets. Building on the concept of virtual trajectories, we provide a Python implementation to generate virtual trajectories from large raw datasets that are typically challenging to process due to their size. We demonstrate the practical utility of these trajectories in assessing speed variability and travel times across different lanes within the INCEPTION dataset.  The virtual trajectory dataset opens future research on traffic waves and their
impact on energy.
 
\end{abstract}


\section{Introduction}

The creation and analysis of vehicle trajectories is foundation in the field of traffic science~\cite{li2020trajectory}. Traffic control, driving safety, infrastructure design, and energy studies all utilize vehicle trajectory data directly or through models built upon trajectory data. Because trajectories contain detailed information about the velocity and acceleration of vehicles in the traffic stream, they are critical to understanding vehicle energy consumption and emissions~\cite{wu2019tracking}. Notable inefficiencies arise in the traffic stream due to bottlenecks, incidents, and oscillatory or stop-and-go driving patterns. The reduction of stop-and-go driving, alone, could yield energy improvements as high as 40\%, along with emissions benefits~\cite{stern2018dissipation, stern2019quantifying}.

Trajectory data sources suitable for creating derivative models or directly analyzing empirical trajectories have been collected in numerous efforts over time, mostly using video data; a few examples include the NGSIM \cite{NGSIM, kovvali2007video}, HighD \cite{krajewski2018highd}, ExiD~\cite{exiDdataset}, and AUTOMATUM~\cite{spannaus2021automatum} on highways and pNEUMA \cite{barmpounakis2020new}, inD \cite{inDdataset}, OpenDD \cite{breuer2020opendd}, Interaction \cite{zhan2019interaction}, and CitySim \cite{zheng2022citysim} on urban surface streets. NGSIM, in particular, has been used to model congestion inefficiency~\cite{treiber2008much, li2014stop}, calibrate car-following models~\cite{chen2010calibration}, study lane-changing behavior~\cite{leclercq2007relaxation}, and much more.

As the size and scale of the systems designed to collect trajectory information continue to expand, they can capture the mechanisms of stop and go driving and assess such traffic patterns~\cite{laval2010mechanism, li2014stop}. But as the scale increases, so does the complexity of the data processing. The I-24 MOTION testbed~\cite{gloudemans202324} is capable of producing vehicle trajectories over a 4.2 mile highway segment on a regular basis, given its fixed infrastructure. The datasets derived from multi-camera systems and sparse camera deployments are susceptible to a range of errors caused by issues with camera synchronization, positional calibration, outages, and occlusions. Computer vision techniques for extracting vehicle trajectories are improving~\cite{gloudemans2023so, wang2022automatic, wang2023online}, which is important for certain microscopic analyses; however, it is also observed that the detailed macroscopic speed fields derived from the trajectories are capable of capturing traffic features such as stop and go waves at a high resolution~\cite{gloudemans202324}. 

This article builds on a recently proposed approach of virtual trajectory generation, introduced in \cite{tsanakas2022generating} and uses the I-24 MOTION INCEPTION  v1.0.0 trajectory dataset as the empirical source data. Virtual trajectories are generated by integrating the velocity of a virtual vehicle in the traffic flow, based on the macroscopic speed field. Given a high-resolution speed field constructed from a suitably detailed dataset, these trajectories exhibit the same traffic patterns as the empirical data, while exhibiting continuous and smooth velocity and continuous acceleration profiles. This noise reduction and robustness to errors compared with empirical trajectories aids in downstream computations such as energy and emissions modeling.

The main contribution of this work is to create a new virtual trajectory dataset called I-24 MOTION INCEPTION VT v1.0,
which has applications in traffic sustainability, modeling, and control. We also provide the intermediate macroscopic speed fields used to generate the virtual trajectories to enable macroscopic analysis. The data and tools will be released when paper is published.
Specifically, the contributions are:
\begin{enumerate}
    \item Create a new virtual trajectory dataset, enabling analysis such as energy modeling that is particularly challenging to conduct on noisy trajectories.
    \item Provide a set of tools with Python implementation to generate virtual trajectories from an empirical trajectory dataset, which is capable of working with raw data that is too large to directly load in memory on many computers (e.g., 70GB). 
    \item Demonstrate the utility of the virtual trajectories, by showing differences in speed variability and travel times experienced by vehicles driving in distinct lanes in the same traffic flow.
\end{enumerate}

\begin{figure*}[!t]
    \centering
    \includegraphics[width=\textwidth]{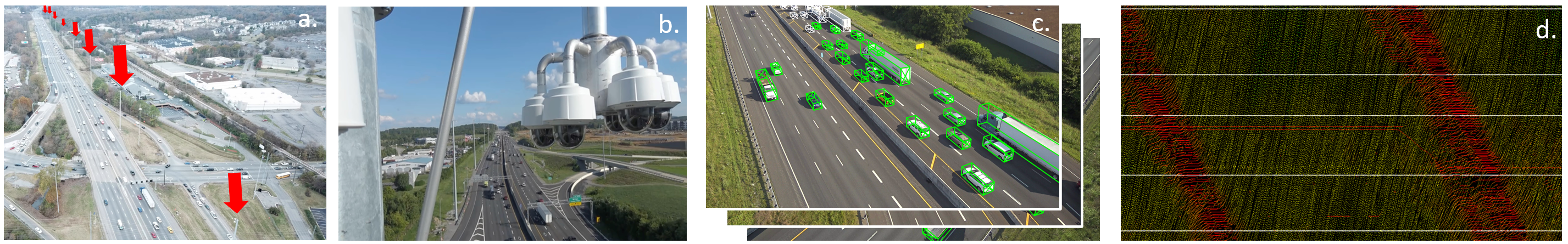}
    \caption{\textbf{I-24 MOTION}: \textbf{a.} Overhead view of a subset of the I-24 MOTION camera poles. \textbf{b.} Close-up of 6 traffic cameras mounted on a single traffic pole. \textbf{c.} Video data is processed with object detection and tracking algorithms to produce 3D positional information for each vehicle at each frame. \textbf{d.} The resulting vehicle positions are stored as vehicle trajectories and are visualized on a time-space diagram. 
    }
    \label{fig:motion}
\end{figure*}

The remainder of this article is organized as follows. In Section \ref{sec:background}, we summarize the I-24 MOTION testbed and the released INCEPTION  v1.0.0 dataset. In Section~\ref{sec:processing_approaches} we describe the core methods we apply from \cite{tsanakas2022generating} to aggregate the trajectory data into a macroscopic speed field, smoothing and interpolation, and generate virtual trajectories. In Section~\ref{sec:results} we describe the constructed virtual trajectories, and demonstrate how they can be used to calculate information that is challenging on the raw dataset. Section~\ref{sec:conclusion} summarizes future directions of research.

\section{Background on I-24 MOTION}
\label{sec:background}

The Interstate 24 MObility Technology Interstate Observation Network (I-24 MOTION) is a new instrument for vehicle trajectory data generation~\cite{gloudemans202324}. It consists of a 4.2 mile stretch of 8-10 lane interstate roadway near Nashville, Tennessee. This portion of roadway is densely covered by 276 4K-resolution traffic cameras mounted on 40 110-foot tall traffic poles, providing a near-seamless viewpoint of the instrumented roadway portion (except at overpasses). All cameras are connected to a centralized compute server via a dedicated fiber-optic network, and video data is processed in parallel with \textit{object detection and tracking algorithms} \cite{gloudemans2021vehicle} to produce vehicle trajectories. Figure \ref{fig:motion} provides a graphic overview of the system.

Vehicle trajectories from the system are output within a coordinate system aligned with the roadway direction of travel \cite{gloudemans2023so} such that the primary coordinate axis is aligned with the primary direction of vehicle travel along the roadway and the secondary axis captures lane information, the standard format for traffic analyses. Each trajectory consists of a fixed vehicle length, width, and height (in feet), a vehicle class (such as \textit{sedan}, \textit{pickup truck}, or \textit{semi truck}) and longitudinal and lateral positional data at fine-grained time intervals (25 Hz). A subsequent smoothing step~\cite{wang2022automatic} is performed on raw trajectories such that speed and acceleration can be directly calculated from the positional data.

The trajectory data produced by I-24 MOTION is large relative to existing vehicle trajectory datasets; however, the current system still has limitations which result in trajectories being \textit{fragmented} (i.e., one vehicle is recorded as several vehicle trajectories as it travels along the roadway). These fragmentations are primarily caused by \textit{i}.) physical object occlusion by either overpasses or by taller vehicles in interior lanes, or \textit{ii}.) failures in the object tracking algorithms. As a result, I-24 MOTION trajectories are not currently suitable for some types of analyses such as long-term vehicle following behavior, travel time analysis, or origin-destination analysis. 

\section{Data processing approach
}
\label{sec:processing_approaches}
The processing approach for generating virtual trajectories from the original trajectory data follows a three-step process proposed in~\cite{tsanakas2022generating}. First, a raw macroscopic speed field, denoted $v_E(t,~x)$, is calculated from the trajectory data using Edie's definition~\cite{edie1963discussion}, for all time $t$ and space $x$. Second, $v_E$  is processed into a smoothed speed field $v_S(t,~x)$ by applying an adaptive smoother \cite{treiber2011reconstructing}. Third, virtual trajectories are generated from the smoothed speed field $v_S$ by integrating virtual vehicles through the speed field \cite{tsanakas2022generating}. 
\subsection{Step 1. Constructing a macroscopic speed field}

Edie~\cite{edie1963discussion} provides an approach to calculate spatiotemporal mean of density, flow speed, and traffic flow from vehicle trajectories.  According to the definition, the density and the flow can be computed from the \textit{total travel time} (TTT) and \textit{total travel distance} (TTD) within an area. Consider a shear box  centered at a point ($t$, $x$), and let   $\Delta t$ and $\Delta x$ denote the height and width of the box. The macroscopic estimates can be computed as:
\begin{align}
    \rho_{E} (t,~x) &= \frac{\text{TTT}(t,~x)}{\Delta x \times \Delta t}, \label{eq:rhoE} \\
    q_{E} (t,~x) &= \frac{\text{TTD}(t,~x)}{\Delta x \times \Delta t},\label{eq:QE} \\
    v_{E} (t,~x) &= \frac{q_{E}(t,~x)}{\rho_{E}(t,~x)} \label{eq:VE},
\end{align}
where $\rho_{E},~q_{E},$ and $v_E$ represent Edie's definition for density, flow and speed respectively. 

\begin{figure}
    \centering
    \includegraphics[width=0.9\linewidth]{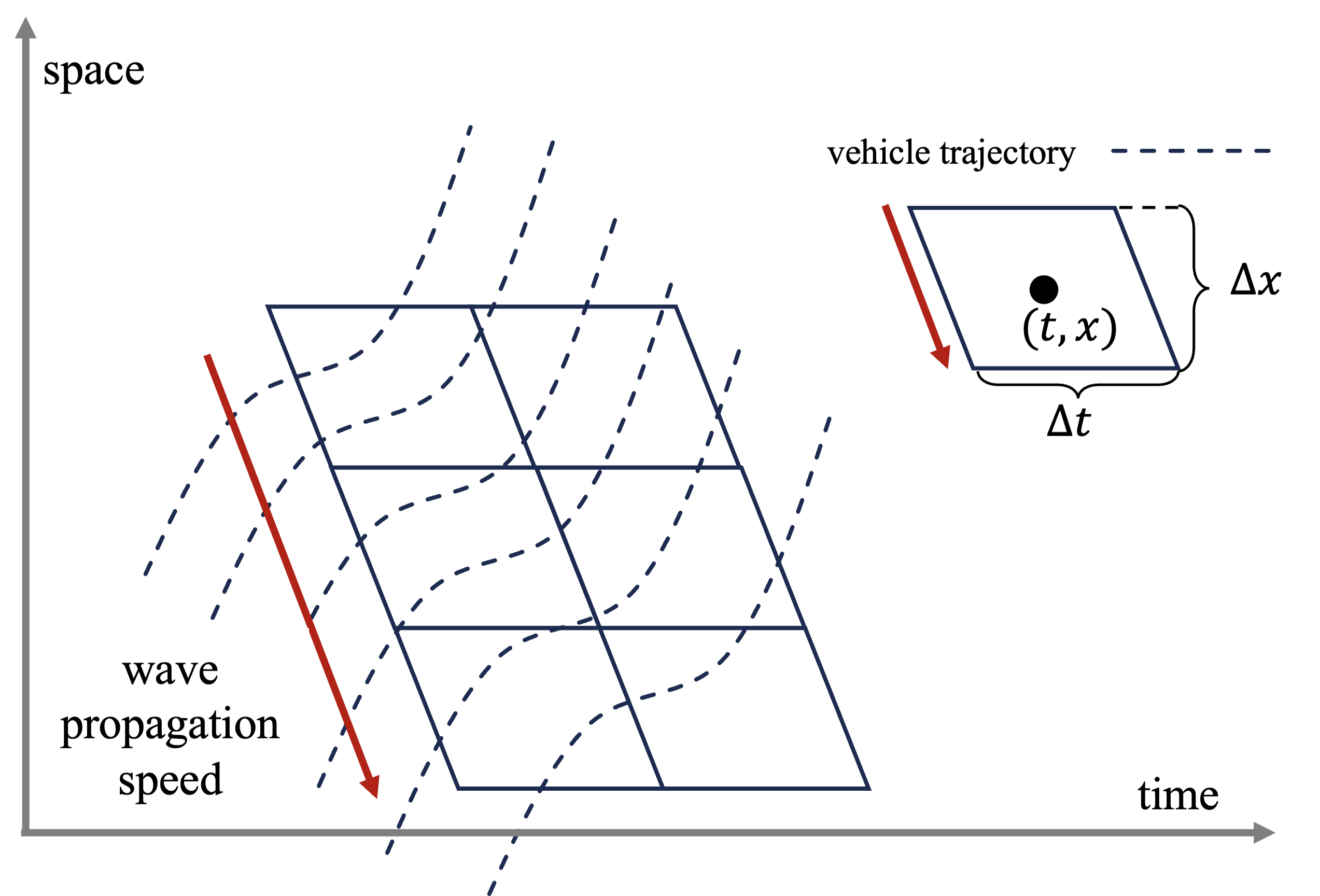}
    \caption{\textbf{Macroscopic speed field calculation}: Illustration of Edie's definition applied to a shear box of size  $\Delta t$ $\times$ $\Delta x$. Dashed lines shows the vehicle trajectories collected from field, and the shear box marks where we quantify the macroscopic measurements. The red arrow points out the stop-and-go wave propagating against the traffic. The shear box's angle matches this wave direction.}
    \label{fig:edie}
\end{figure}

Rather than assuming a rectangle for the area, we apply the shear box approach. As highlighted in the study by \cite{he2017constructing}, the use of shear grid cells has improved performance over traditional rectangular grid cells in mean speed measurements. This is particularly evident in the context of traffic waves, where shear cells align more closely with the backward-moving wave direction (illustrated in Figure~\ref{fig:edie}), thus maintaining more homogeneous traffic conditions within the area. For this work we focus specifically on the macroscopic speed. 
 
\subsection{Step 2. Constructing a smoothed speed field} The \textit{Adaptive smoothing method} (ASM) developed by 
\cite{treiber2011reconstructing} is a frequently used smoothing and interpolation algorithm to construct a continuous spatiotemporal mean speed field. While it is particularly useful for data from fixed infrastructure sensors such as radar units or inductive loops, it is also applicable to the macroscopic speed data generated from the I-24 MOTION trajectories. This can be particularlly helpful, for example, to interpolate speeds in areas that are occluded from view, such as under bridges.

The main idea of ASM lies in separating the mean speed field into two entities: a free-flow field and a congested field. Following this division, the method exploits the distinct and regular information propagation velocities in free-flow and congested traffic. Succinctly, ASM smooths and interpolates data in free-flow along lines corresponding to the free-flow speed of traffic, and smooths and interpolates data in congestion along lines corresponding to the backward propagating wave speed. For the complete mathematical description of the smoother, refer to  \cite{treiber2011reconstructing}.%

\subsection{Step 3. Virtual trajectory generation}
A standard approach to generate trajectories from a macroscopic speed field is to calculate the position $p(t)$ of a vehicle assuming the velocity dynamics of the vehicle are computed as follows:
\begin{equation}
\frac{\mathrm{d}p(t)}{\mathrm{d}t} = v_S(t,p(t)),
\label{eq:euler}
\end{equation}
given an initial condition $p(0)=p_0$. The solution to the ordinary differential equation \eqref{eq:euler} can be approximated with a forward Euler method with a small timestep.

In the case where the integration timestep used in the ordinary differential equation is small relative to the width $\Delta t$ used to generate the macroscopic speed field, the resulting trajectories may have quantization artifacts. Cubic interpolation \cite{fritsch1980monotone} is suggested by~\cite{tsanakas2022generating} to improve the regularity of the resulting trajectories, which we also adopt.

\subsection{Implementation} The main challenge to implement virtual trajectories on I-24 MOTION data is due to the massive size of the dataset. Notably, the size of the raw trajectory data for a single 4 hour time-space diagram is approximately 70GB uncompressed, and typically contains around 100 to 310 million trajectory points. 

Since the raw data is too large to load into memory on many computers, care is needed on the implementation. For example, we process the raw noisy trajectory data sequentially, incrementing the total travel time and total distance in each box in which the trajectory is observed. This allows us to avoid needing to load all trajectories into memory. Compared to a calculation in which we process all trajectories within a given box before moving to the next box, we also avoid the need to repeatedly load trajectories. This approach retains the ability to  parallelize the computations, further speeding up the calculations.

As part of this work, we provide our Python implementation detailing step-by-step how to access the data, implement the methods, and generate the results. The Python code will be made public at~\cite{I24MotionURL}.

\section{Results 
}
\label{sec:results}

In this section, we summarize the virtual trajectory creation and analysis during a typical morning commute (Tuesday, November 22, 2022) on I-24 W. First, we create the raw mean speed field derived from the empirical trajectory data. We then calculate the smoothed mean speed field obtained through ASM smoothing, and construct virtual trajectories in each lane. Finally, we illustrate the utility of the virtual trajectories through an exploration of the travel times through stop and go waves in different lanes of travel.

\subsection{Raw macroscopic speed field from trajectory data}

\begin{figure*}
    \centering
    \includegraphics[width=\textwidth]{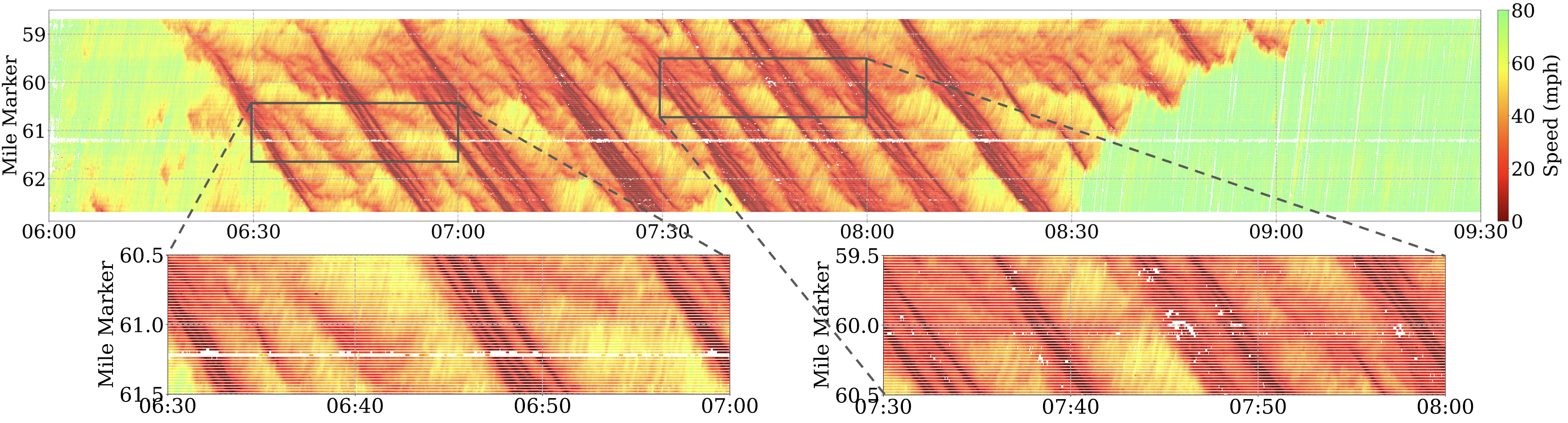}
    \caption{\textbf{Raw mean speed field}: This time-space diagram showcases the raw speed field computed using Equation~\eqref{eq:VE} for the HOV (left-most lane), obtained from one day of the I-24 MOTION INCEPTION  dataset. Upon closer inspection in the insets, instances of missing data and outliers are observed within the field. }
    \label{fig:speed_field}
\end{figure*}

The speed field for a single lane generated from trajectory data using Edie's definition is shown in Figure~\ref{fig:speed_field}. This speed field corresponds to the leftmost lane, which is a \textit{high occupancy vehicle }(HOV) lane. Traffic flows in the direction of decreasing mile markers. 

With the selected discretization of ($\Delta x = 0.02$ miles and $\Delta t = 4$ seconds), the details of the traffic characteristics are preserved.  It is evident that the congested period of the traffic stream consists of prominent congestion waves (parallel red streaks in the diagram), which move upstream (against the flow of traffic) at a relatively consistent speed. The stop-and-go traffic pattern emerges around 6:20 AM, persists until 9:00 AM, and is followed by a return to free-flow conditions. This level of granularity in the speed field enables the observation of other nuanced traffic waves properties such as bifurcation patterns, growth in width, and derivative waves, which are of interest for further analysis.

\begin{figure*}
    \centering
    \includegraphics[width=0.9\textwidth]{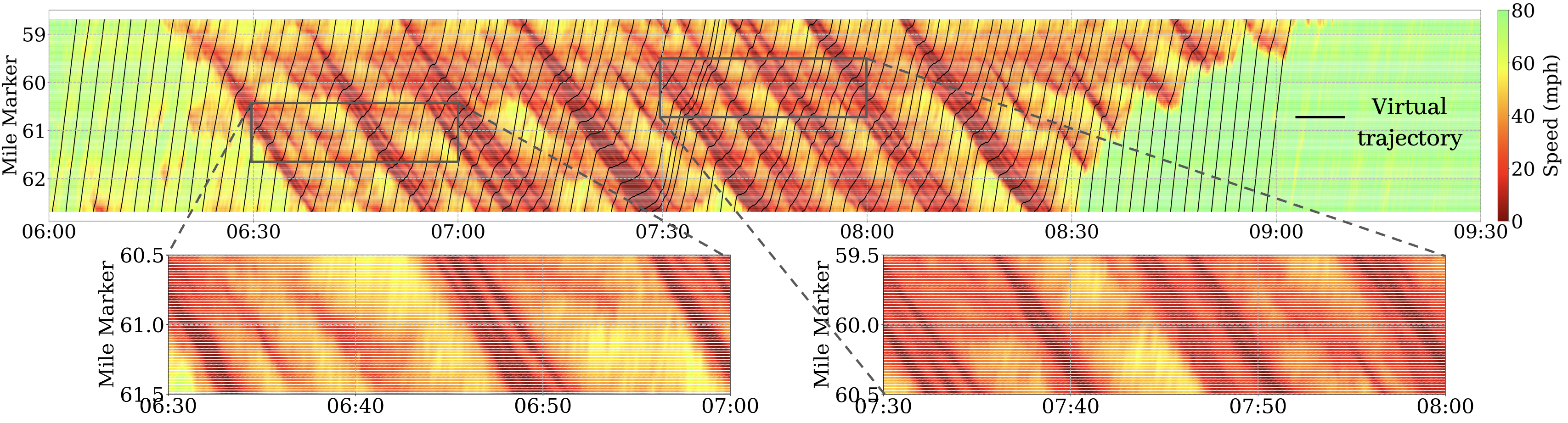}
    \caption{\textbf{Smoothed speed field and samples of virtual trajectories}: This time-space diagram illustrates the smoothed speed field computed using the ASM method for a specific lane (HOV lane) derived from one day of the I-24 MOTION INCEPTION  dataset. The speed field appears smooth and fully reconstructed. The virtual trajectories shown here (black lines) are captured at departure intervals of 120 seconds (2 minutes) for illustration.}
    \label{fig:speed_field_smooth}
\end{figure*}
\subsection{Smoothed mean speed field}

The raw mean speed field (Figure~\ref{fig:speed_field} exhibits imperfections and missing data (notably at mile marker 61.2) due to artifacts in the source trajectory data, such as offline cameras and obstructed fields of view. Consequently, these artifacts create gaps in local speed data that hinder the generation of continuous virtual trajectories and necessitate further smoothing and data imputation. We apply the smoothing method detailed in Section~\ref{sec:processing_approaches} to the raw mean speed field derived from Edie's definition.

The outcome following ASM is shown in Figure~\ref{fig:speed_field_smooth}. The smoothing method successfully imputes all gaps present in the raw mean speed field. Notably, the intricate structures of the traffic waves remain preserved through this process. This complete speed field now enables the computation of virtual trajectories for any chosen departure time using the Euler forward update method.

A series of virtual trajectories, depicted as black lines with departures spaced at a consistent 2-minute interval, are overlaid onto the smoothed mean speed field in Figure~\ref{fig:speed_field_smooth}. It is apparent that the travel time varies based on the chosen departure time and the resulting number and severity of traffic waves that are encountered.

\subsection{Lane comparison using virtual trajectories}

\begin{figure}
    \centering
    \includegraphics[width=\linewidth]{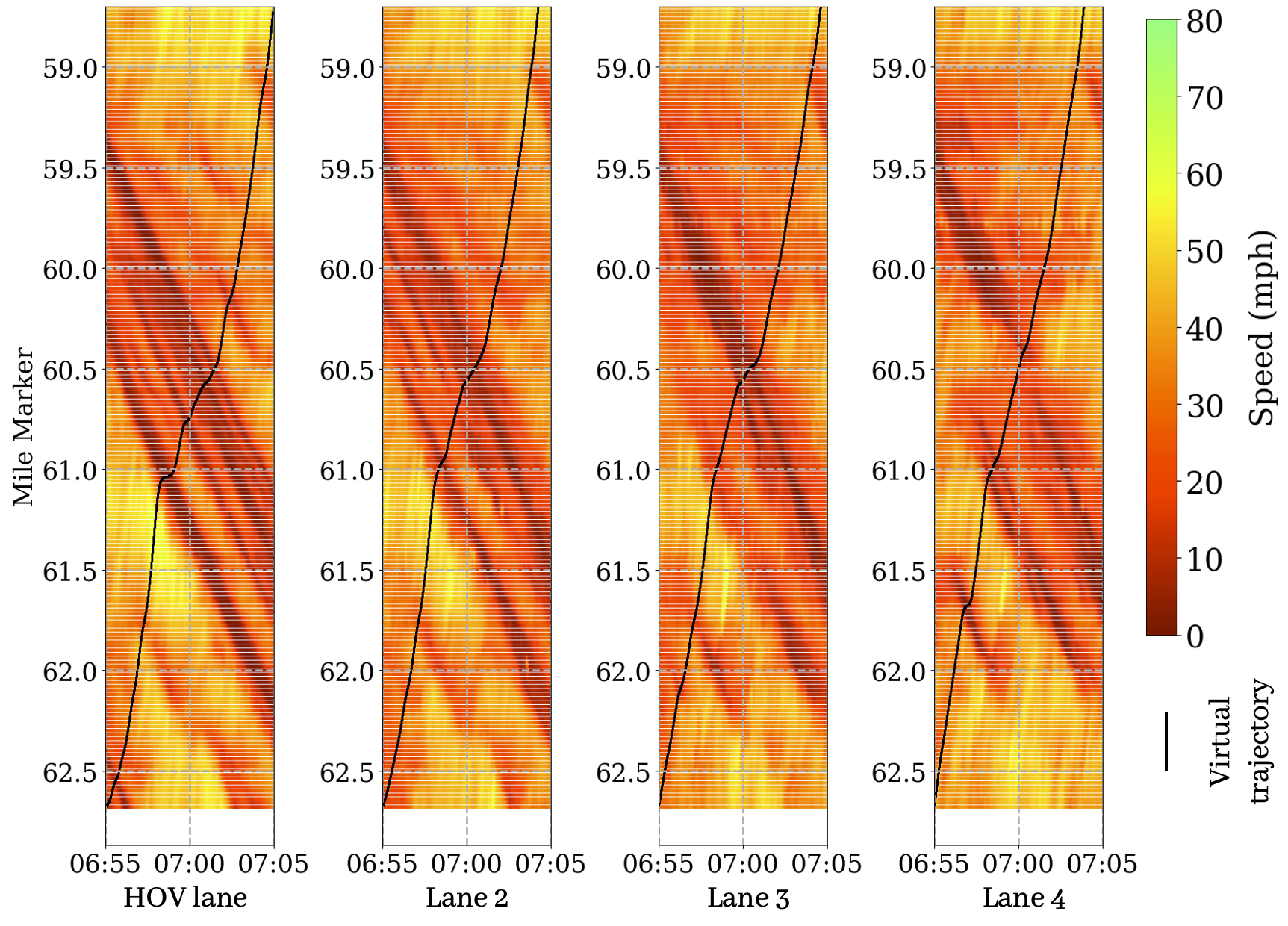}
    \caption{\textbf{Lane comparison:} Smoothed mean speed field for ten minutes in each lane. A sample virtual trajectory for each lane is shown in black line, with identical departure time and location across the lanes.}
    \label{fig:vt_case}
\end{figure}

Figure~\ref{fig:vt_case} illustrates a virtual trajectory from each of the four lanes, all originating from the identical position (mile marker 62.7) and departure time (6:55 AM). Despite the synchronization in departure time and location, distinct wave patterns emerge across the lanes. Traffic waves in the HOV lane appear more severe and have a concentrated impact on the trajectory; waves encountered in other lanes are present in a similar pattern, but with lower impact allowing for less variation in speed. 

\begin{figure}
    \centering
    \includegraphics[width=\linewidth]{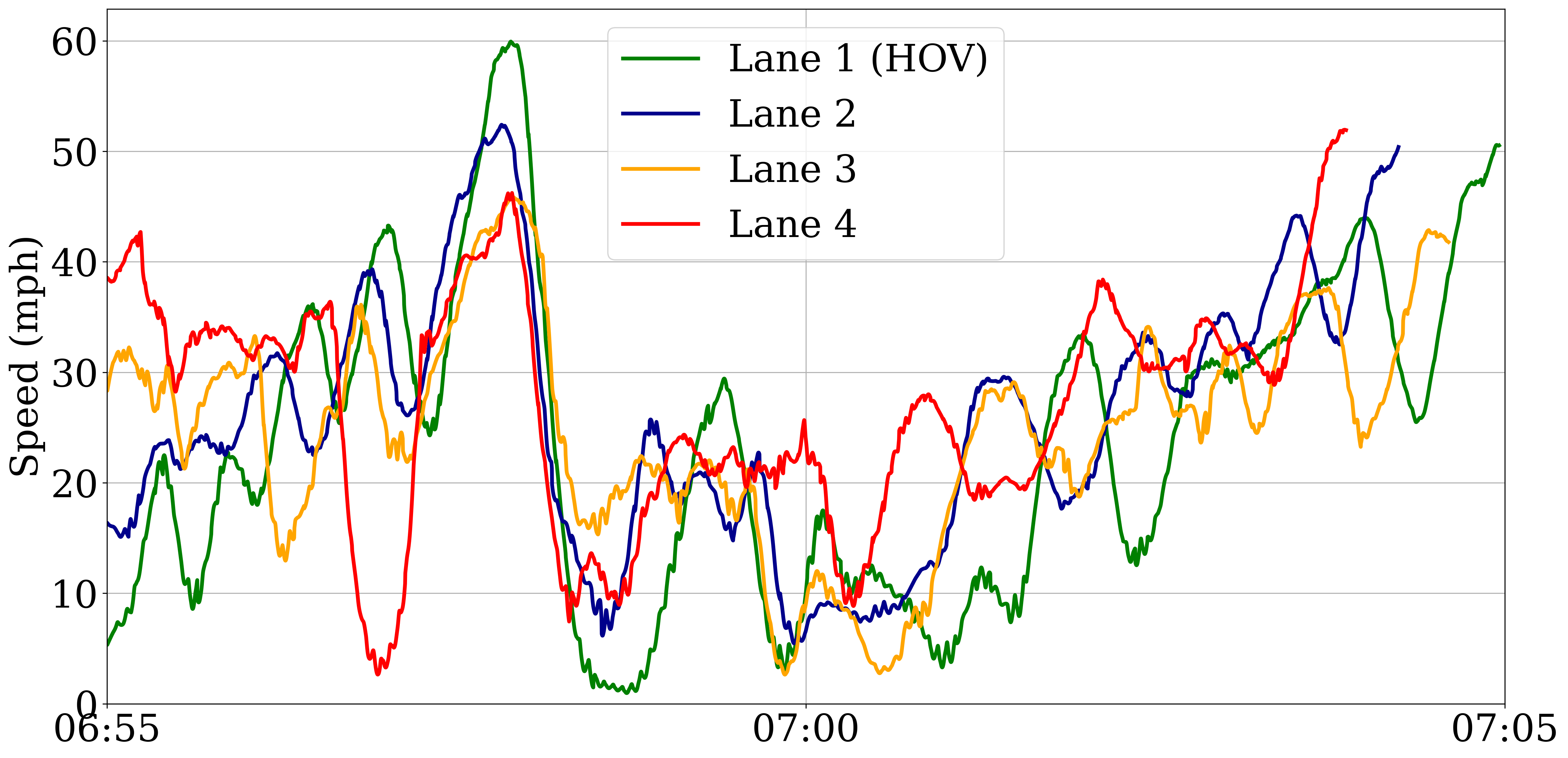}
    \caption{\textbf{Speed time-series}: four speed profiles for each virtual trajectory generated from each lane starting at the same time and position.}
    \label{fig:vt_case_speed}
\end{figure}

Figure~\ref{fig:vt_case_speed} shows the speed time-series for each of the four virtual trajectories in Figure~\ref{fig:vt_case}. The difference in wave patterns across lanes is reflected in the timing of their speed maxima and minima, as well as the ultimate travel times of the virtual trajectories. In this selected time interval, the travel duration in lane 4 is nearly 2 minutes faster than that in the HOV lane. Furthermore, the speed fluctuations in the HOV lane exceed those in other lanes, peaking at 60 mph while other lanes reach only 50 mph and nearly reaching 0 mph at multiple points. At 6:58 AM, the speed in the HOV lane drops from 60 mph to near standstill within 40 seconds, a contrast to a lower slowdown range observed in other lanes. Additionally, while there is a general synchronization in speed across lanes, at a smaller scale, these fluctuations are slightly out of phase (e.g., during the overall speed increase period from 7:00 to 7:05 AM).

We generated 713 virtual trajectories per lane (total of 2,852 virtual trajectories), departing at 15-second intervals between 6:00 to 9:00 AM, and computed mean and standard deviation of travel time, along with the standard deviation of speed for each trajectory. The speed standard deviation serves as a proxy for fuel consumption \cite{barth2009energy}. The findings are summarized in Table~\ref{tab:more}. Notably, within this timeframe, the HOV lane exhibits the shortest mean travel time (471 seconds), while demonstrating the highest speed standard deviation (14.87 mph). This observation broadly aligns with the speedup and slowdown characteristics evident in Figure~\ref{fig:vt_case_speed}, further highlighting the dramatic speed fluctuations within the HOV lane.

\begin{figure}
    \centering
    \includegraphics[width=0.9\linewidth]{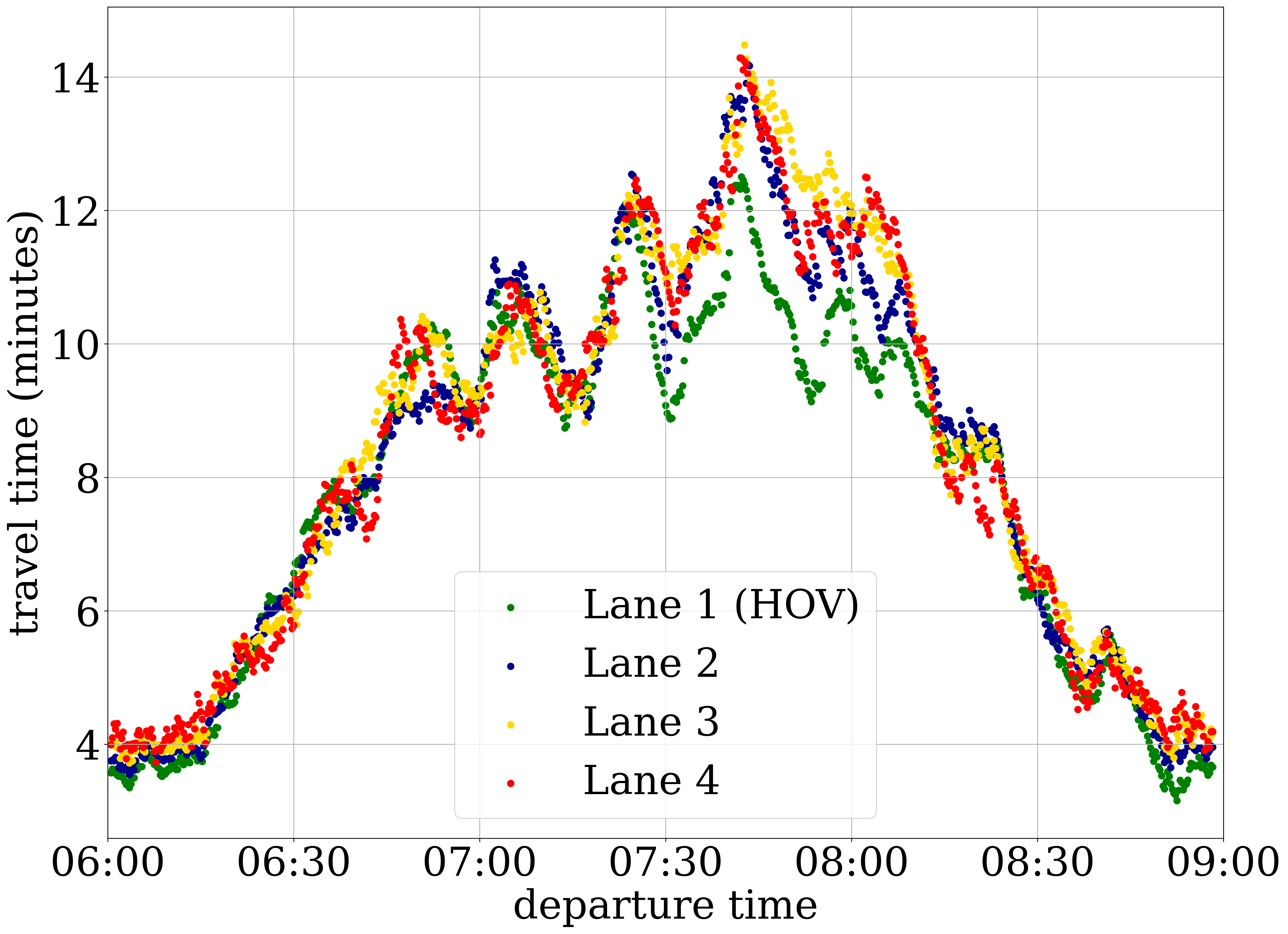}
    \caption{\textbf{Travel time with different departure time}: the x-axis is departure time of the virtual trajectories and y-axis is the travel time in minute of each virtual trajectory. The ``departure time" here refers to the initial time of virtual trajectory, in the context of real traffic, it refers to the time vehicle enters this stretch of highway.}
    \label{fig:statistics}
\end{figure}

Figure~\ref{fig:statistics} illustrates the relationship between departure time and travel time for all 2,852 virtual trajectories across all lanes. During periods of free flow (roughly before 6:15 AM and after 8:45 AM), the travel time across all lanes stabilizes at approximately 4 minutes with expected stratification across lanes. However, as congestion sets in, travel times begin to increase, peaking around 7:45 AM. During heavily congested periods, the high occupancy vehicle lane demonstrates travel times that are 1-2 minutes shorter compared to other lanes. An interesting trend emerges across all lanes: during congestion, departing just 10 minutes later or earlier could lead to a travel time variation of up to 4 minutes, highlighting the sensitivity of travel times to departure timings over this specific stretch of roadway due to the traffic waves.

\begin{table}
\centering
\caption{Statistics for the virtual trajectories (sample size N = 713 for each lane) collected across different lanes}
\begin{tabular}{@{}ccccc@{}} 
\toprule 
\textbf{} & \textbf{HOV} & \textbf{Lane 2} & \textbf{Lane 3} & \textbf{Lane 4} \\
\midrule 
mean travel time (min) & 7.86 & 8.26 & 8.42 & 8.33 \\
st.d. travel time (min) & 2.63 & 2.94 & 3.01 & 2.93 \\
mean speed st.d. (mph) & 14.87 & 13.39 & 12.06 & 11.50 \\
\bottomrule 
\end{tabular}
\label{tab:more}
\end{table}

\section{Conclusions}
\label{sec:conclusion}

This work generates virtual trajectories for a large scale dataset known as I-24 MOTION INCEPTION . While the dataset captures the macroscopic pattern of traffic waves in high fidelity, fragments and other errors in the raw trajectories inhibit its application to specific microscopic analysis, such as energy estimates. To overcome this limitation, we compute virtual trajectories using a standardized approach, and provide the Python implementation and resulting trajectories with this work. This lowers the degree of effort required for researchers to work with I-24 MOTION data, which can otherwise be challenging due to the overall data size. This opens future research on traffic waves and their impact on a range of topics ranging from safety to energy consumption. %

\section*{Acknowledgement}
The authors wish to express their gratitude to $\mathrm{M.~Treiber}$ for insights on the dimensions of Edie's box, and $\mathrm{B.~Piccoli}$, $\mathrm{J.~Sprinkle}$, and $\mathrm{M.~Nice}$ for discussions on data analysis and virtual trajectories. This work is supported by the National Science Foundation (NSF) under Grant No. 2135579 and the NSF Graduate Research Fellowship Grant No. DGE-1937963 (Gloudemans). This material is based upon work supported by the U.S. Department of Energy’s Office of Energy Efficiency and Renewable Energy (EERE) award number CID DE-EE0008872. The views expressed herein do not necessarily represent the views of the U.S. Department of Energy or the United States Government.


\bibliographystyle{IEEEtran}
\bibliography{bib}

\end{document}